\documentclass[reprint,prl,showpacs,preprintnumbers,amsmath,amssymb,twocolumns,footinbib]{revtex4-1}

\usepackage{graphicx}
\usepackage{epstopdf}
\usepackage{dcolumn}
\usepackage{bm}

\begin{document}

\title{Theory of microwave-assisted supercurrent in quantum point contacts}

\author{F.S. Bergeret$^1$, P. Virtanen$^{2,3}$, T.T. Heikkil\"{a}$^{2}$,
J.C. Cuevas$^{4}$}

\affiliation{$^{1}$Centro de F\'{\i}sica de Materiales (CFM-MPC), Centro Mixto 
CSIC-UPV/EHU and Donostia International Physics Center (DIPC),
Manuel de Lardizbal 5, E-20018 San Sebasti\'an, Spain.\\
$^{2}$Low Temperature Laboratory, Aalto University School of Science and Technology, 
P.O. Box 15100, FI-00076 AALTO, Finland.\\
$^{3}$Institute for Theoretical Physics and Astrophysics,
University of W\"urzburg, D-97074 W\"urzburg, Germany.\\
$^{4}$Departamento de F\'{\i}sica Te\'orica de la Materia Condensada,
Universidad Aut\'onoma de Madrid, E-28049 Madrid, Spain.}

\date{\today}

\begin{abstract}
We present a microscopic theory of the effect of a microwave field on the 
supercurrent through a quantum point contact of arbitrary transmission. Our
theory predicts that: (i) for low temperatures and weak fields, the 
supercurrent is suppressed at certain values of the superconducting phase,
(ii) at strong fields, the current-phase relation is strongly modified
and the current can even reverse its sign, and (iii) at finite temperatures,
the microwave field can enhance the critical current of the junction. Apart 
from their fundamental interest, our findings are also important for the 
description of experiments that aim at the manipulation of the quantum state of 
atomic point contacts.
\end{abstract}


\maketitle
\begin{figure}[b]
\begin{center}
\includegraphics[width=.75\columnwidth]{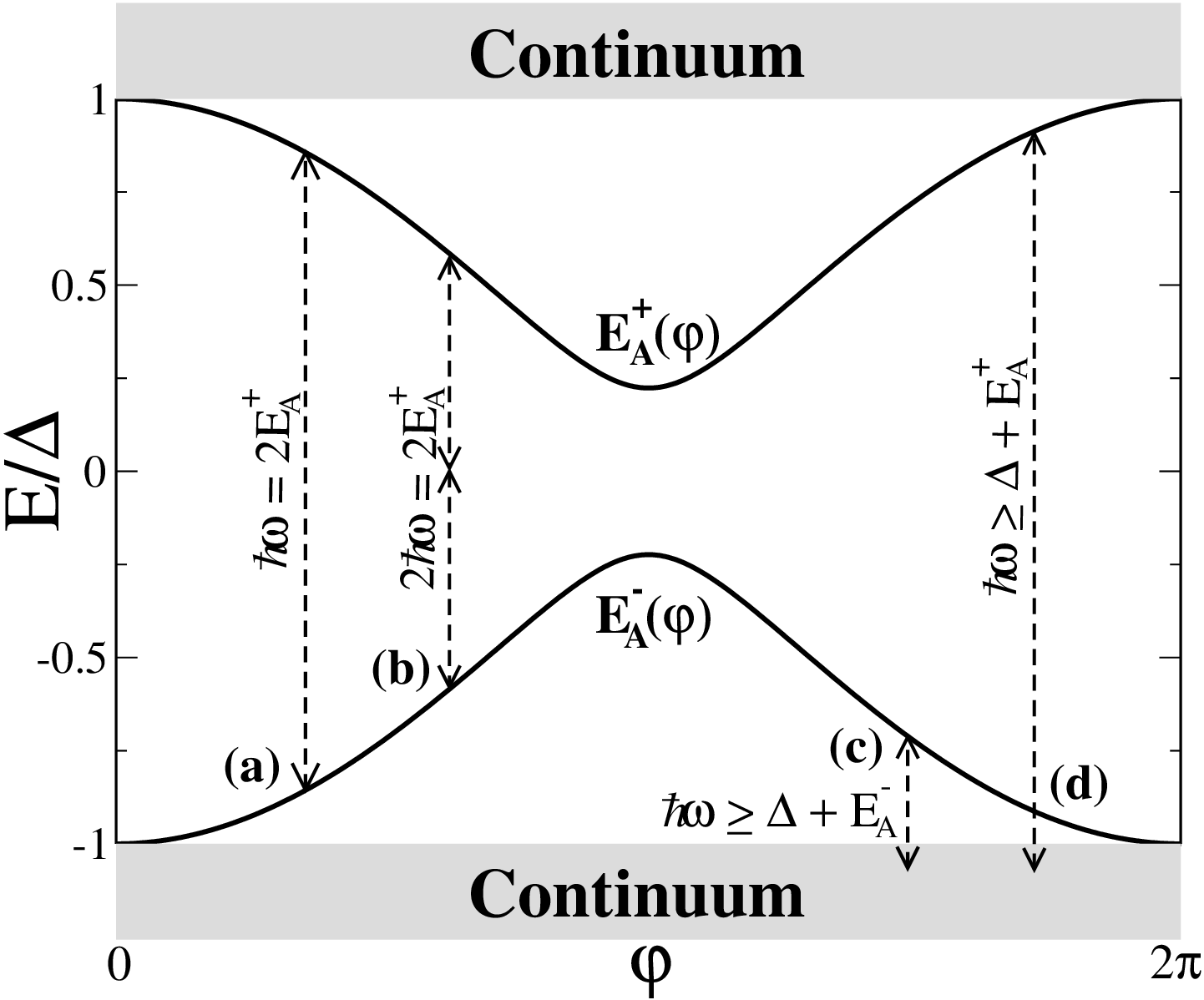}
\caption{Dispersion relation of the ABSs in a single-channel QPC with
transmission $\tau = 0.95$. The vertical dashed lines indicate 
one- and two-photon transitions between
the ABSs (a-b), and transitions between the continuum and ABSs (c-d). We have chosen $E=0$ at the Fermi energy.
% some relevant
%microwave-induced transitions: (a-b) one- and two-photon transitions between
%the ABSs, (c) transition from the continuum to the lower ABS, and (d)
%transition from the continuum to the higher ABS.
} \label{model}
\end{center}
\end{figure}

The DC Josephson effect is one of the most striking 
examples of macroscopic quantum coherence. In the context of superconductivity 
this phenomenon manifests as the flow of a dissipationless DC current through a 
junction in the absence of any voltage \cite{Josephson1962,Golubov2004}. Since 
its discovery in the early 1960s, this effect has been observed in a large 
variety of weak links such as tunnel junctions \cite{Barone1982}, microbridges 
\cite{Likharev1979}, atomic contacts \cite{Koops1996}, carbon nanotubes 
\cite{Kasumov1999}, semiconductor nanowires \cite{Doh2005} and graphene 
\cite{Heersche2007}. In spite of their intrinsic differences, the DC Josephson
effect in these systems can be described in a unified manner. It has been 
shown that for constrictions shorter than the superconducting coherent length, 
the Josephson current is carried exclusively by a single pair of \emph{Andreev 
bound states} (ABSs) \cite{Furusaki1991,Beenakker1991}. In the simplest case 
of a single-channel contact of transmission $\tau$, these states 
appear at energies $E^{\pm}_{\rm A}(\varphi,\tau) = \pm \Delta [ 1 - \tau \sin^2
(\varphi/2) ]^{1/2}$, where $\Delta$ is the superconducting gap and $\varphi$ 
is the phase difference between the order parameters on both sides (see 
Fig.~\ref{model}). These states carry opposite supercurrents $I^{\pm}_{\rm A}
(\varphi) = (2e/\hbar) \partial E^{\pm}_{\rm A}/ \partial \varphi$, which are 
weighted by the  occupation of the ABSs. A more complex weak link, like 
the ones mentioned above, can be viewed as a collection of independent conduction
channels, characterized by a set of transmission coefficients. The 
 supercurrent through it is  given by the sum of the contributions 
from the individual channels \cite{Beenakker1991}.

This unified microscopic picture of the DC Josephson effect has been
recently confirmed experimentally in the context of atomic contacts
\cite{Rocca2007}, where the current-phase relationship has been
directly measured. These experiments mainly probed the ground Andreev
state and, as it is stated in Ref.~[\onlinecite{Rocca2007}], it would
be of great interest to probe also the excited state, for instance,
through microwave spectroscopy. This leads us to the central question
of the present work: How does a microwave radiation modify the
supercurrent of a single-channel quantum point contact (QPC)? Apart
from its fundamental interest for the field of mesoscopic
superconductivity, this question is also of great relevance for the
proposals of using the ABSs of a QPC as the two states of a quantum bit
\cite{Desposito2001,Zazunov2003,Zazunov2005}, whose quantum state can
be probed by means of current measurements. 

Surprisingly, there is no complete answer to the question posed above. The
theoretical analysis of the microwave-assisted supercurrent in point contacts
has either been addressed within phenomenological approximations
or in the limiting case of very weak fields \cite{Shumeiko1993,Gorelik1995}.
In this letter, we present a microscopic theory of the effect of a microwave
field on the supercurrent of a single-channel quantum point contact valid for
arbitrary range of parameters. Our theory based on the Keldysh technique
predicts the following novel effects: (i) at low temperatures, the supercurrent
can be strongly suppressed at certain values of the phase due to resonant 
microwave-induced transitions between the two ABSs (processes of type \emph{a} and 
\emph{b} in Fig.~\ref{model}). (ii) As the radiation power increases, the supercurrent-phase 
relation is strongly modified and it can even reverse its sign. (iii) At finite
temperatures, the radiation can induce the transition of quasiparticles from 
the continuum to the lower ABS leading to an enhancement of the critical current
as compared to the case in the absence of microwaves (process of type \emph{c} in 
Fig~\ref{model}). We also compare our results to a two-level model (TLM), where 
the QPC is described exclusively in terms of the ABSs and show that effects (ii) 
and (iii) fall out of the scope of TLM. This is especially relevant for the 
quantum computing applications. 

We consider a QPC consisting of two identical 
superconducting electrodes (denoted as $L$ and $R$) and linked by a single 
conduction channel of transmission $\tau$. Our goal is to compute the supercurrent 
through this QPC when it is subjected to a monochromatic microwave field of frequency 
$\omega$. We assume that the external radiation generates a time-dependent voltage 
$V(t) = V_0 \sin \omega t$ \cite{Barone1982}. According to the Josephson relation, 
this voltage induces a time-dependent superconducting phase difference given by
$\phi(t) = \varphi + 2 \alpha \cos \omega t$, where $\varphi$ is the DC part
of the phase and $\alpha = eV_0/\hbar \omega$ is a parameter that measures the
strength of the coupling to the electromagnetic field and that is proportional
to the square root of the radiation power at the junction. Following 
Refs.~\cite{Zaitsev1998,Nazarov1999}, the current through a QPC with
an arbitrary time-dependent voltage can be computed as $I(t) = (e/4\hbar) {\rm Tr}
\hat \tau_3 \hat I^K(t,t)$, where $\hat \tau_3$ is the third Pauli matrix and 
$\hat I^K(t,t)$ is the Keldysh component of the current matrix given by
\begin{equation}
\check I(t,t^{\prime}) = 2 \tau \left[\check G_L , \check G_R \right]_{\circ}
\circ \left[ 4 - \tau \left(2- \left\{ \check G_L , \check G_R \right\}_{\circ}
\right) \right]^{-1} (t,t^{\prime})  \label{matrixJ}.
\end{equation}
Here the symbol $\check{\,}$ represents $4\times4$ matrices in Keldysh-Nambu space
and the symbol $\circ$ denotes the convolution over intermediate time arguments.
Moreover, $\check G_{L(R)}$ are the quasiclassical Green functions for the left 
and right electrodes, which can be expressed as $\check G_j(t,t^{\prime}) = 
e^{i\phi_j(t) \hat \tau_3/2} \check g_j(t-t^{\prime}) e^{-i\phi_j(t^{\prime})
\tau_3/2}$. Here, $\check g(t)= \int (dE/2\pi) e^{-iE t/\hbar} \check g(E)$ is the 
equilibrium Green function of the leads and $\phi_j(t)$ is the time-dependent 
phase of the $j$ superconductor, $j=L,R$, i.e.\ $\phi_L(t) = -\phi_R(t) = \phi(t)/2$. 
The retarded ($R$), advanced ($A$) and Keldysh ($K$) components of $\check g(E)$ 
adopt the form: $\hat g^{R(A)}(E) = g^{R(A)}(E) \hat \tau_3 + f^{R(A)}(E) i 
\hat \tau_2$ and $\hat g^{K}(E) = \left[\hat g^{R}(E) - \hat g^{A}(E)\right] 
\tanh(E/2k_{\rm B}T)$, where $g^{R(A)}(E) = E/[(E \pm i\eta)^2-\Delta^2]^{1/2}$, $f^{R(A)}(E) = \Delta/[(E \pm i \eta)^2-\Delta^2]^{1/2}$, and $\eta\rightarrow 0^+$

It is easy to show that, due to the time dependence of the phase, the lead
Green functions $\check G_{L(R)}$, and any product of them, admit the following
Fourier expansion $\check G(t,t^{\prime}) = \sum^{\infty}_{m=-\infty} e^{im\omega t^{\prime}} \int \frac{dE}{2\pi} e^{-i E(t-t^{\prime})/\hbar} \check G_{0m}(E)$, 
where $\check G_{nm}(E) \equiv \check G(E+n\hbar \omega, E+m\hbar \omega)$
are the corresponding Fourier components in energy space. Thus, $\check I$ in
Eq.~(\ref{matrixJ}) can be written as a product of matrices in energy space. 
In particular, its Keldysh component is given by
\begin{equation}
\hat I_{nm}^K = \sum_l[\hat A_{nl}^R \hat X_{lm}^K + \hat A_{nl}^K \hat X_{lm}^A] .
\label{eq-J}
\end{equation}
Here, we have defined the 
matrices $\check A_{nm} \equiv 2 \tau [\check G_L,\check G_R]_{nm}$ and 
$\check X_{nm} = [4\check 1-\tau(2-\{\check G_L,\check G_R\})]^{-1}_{nm}$,
which can be  determined from the Fourier components of $\check G_{L(R)}$.
Once the components of $\hat I^K$ are obtained from Eq.~(\ref{eq-J}), one can 
compute the current. We are only interested in the DC component, which reads
\begin{equation}
I(\varphi,\omega,\alpha) = \frac{e}{4\hbar} \int \frac{dE}{2\pi} {\rm Tr} 
\hat \tau_3 \hat I^K_{00}(E,\varphi,\omega,\alpha) .
\label{Idc}
\end{equation} 
The DC current can only be calculated analytically in certain limiting cases
like in the absence of microwaves, in the tunnel regime or for very weak fields.
In general, Eq.~(\ref{eq-J}) and the current have to be evaluated numerically.

In the absence of microwaves,  the current  from Eq. (\ref{Idc})  can be written as a sum
of the contributions of the two ABSs as $I_{\rm eq}(\varphi) = I^-_{\rm A} 
n_{\rm F}(E^-_{\rm A}) + I^+_{\rm A} n_{\rm F}(E^+_{\rm A})$, $n_{\rm F}(E)$
being the Fermi distribution function, which yields
\begin{equation}
I_{\rm eq}(\varphi) = \frac{e\Delta^2}{2\hbar} \frac{\tau \sin\varphi}
{E^+_{\rm A}(\varphi)} \tanh \left( \frac{E^+_{\rm A}(\varphi)}
{2k_{\rm B}T} \right) .\label{Ieq}
\end{equation}
In equilibrium, the maximum current is then obtained at temperature $T=0$,
when the lower ABS is fully occupied and the upper one is empty.  In the presence of 
the microwave field the simplest approach is the so-called adiabatic approximation \cite{Barone1982}, which consists in replacing   the stationary phase 
$\varphi$ in Eq.~(\ref{Ieq}) by the time-dependent phase $\phi(t)$. This leads to 
\begin{equation}
I_{\rm ad}(\varphi,\alpha) = \sum_{n=1}^{\infty}I_n J_0(2n\alpha) 
\sin(n\varphi), \label{Iad}
\end{equation}
where $I_n=(1/\pi) \int_0^{2\pi} d\varphi \; I_{\rm eq}(\varphi) \sin(n\varphi)$
are the harmonics of the equilibrium current-phase relation and $J_0$ is
the zero-order Bessel function of the first kind. 

\begin{figure}[t]
\begin{center}
\includegraphics[width=1\columnwidth]{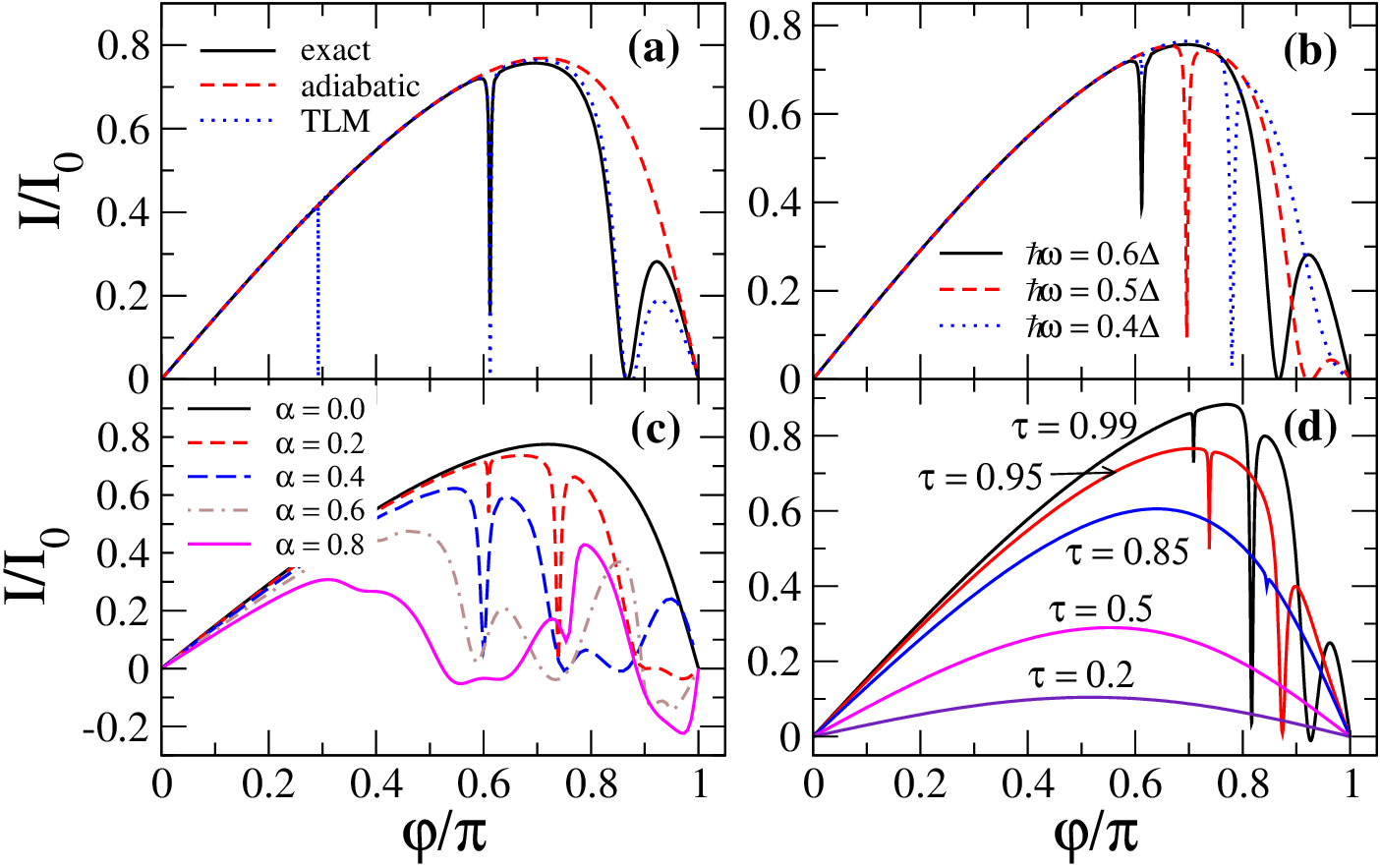}
\caption{(Color online) Zero-temperature supercurrent, in units of $I_0=e\Delta_0/\hbar$, 
as a function of the phase. (a) The solid line shows the exact result for $\tau=0.95$,
$\hbar \omega = 0.6 \Delta$ and $\alpha=0.1$. The dashed line corresponds to the
approximation of Eq.~(\ref{Iad}) and the dotted line to the TLM. Rest of the panels show the exact result for  (b)  $\tau=0.95$, $\alpha=0.1$, and different frequencies,  (c)  $\tau=0.95$, $\hbar \omega = 0.3 \Delta$, and different  $\alpha$'s,  (d)  $\hbar \omega = 0.3 \Delta$, $\alpha=0.1$ and different $\tau$'s.
\label{CPR}}
\end{center}
\end{figure}

In Fig.~\ref{CPR}(a) we show the zero-temperature current-phase relation (CPR)
computed numerically from Eq.~(\ref{Idc}) for a highly transmissive channel with
$\tau=0.95$ and  a weak field $\alpha=0.1$ of frequency $\hbar \omega = 0.6 
\Delta$.  For comparison we also show as a dashed line the result obtained with
the adiabatic approximation of Eq.~(\ref{Iad}). The main difference is that the exact result shows  
a series of dips where the current  is largely suppressed. These dips originate from microwave-induced transitions
from the lower ABS to the upper one that enhance the population of the latter, 
diminishing the supercurrent (see processes \emph{a} and \emph{b} in Fig.~\ref{model}). 
Such transitions can occur whenever the Andreev gap (distance between the ABSs) 
is equal to a multiple of the microwave frequency, i.e.\ $2E^+_{\rm A}(\varphi) 
= n \hbar \omega$, where $n = 1,2,...$ can be interpreted as the number of 
photons involved in the transition. For small  values of $\alpha$ the ABSs  remain almost unchanged, thus the resonant processes 
take place at phases given by 
\begin{equation}
\varphi_n = 2 {\rm arcsin} \sqrt{[1-(n\hbar\omega/2\Delta)^2]/\tau},
\; \; n= 1,2,\dots \label{phin}
\end{equation}
This expression reproduces accurately the positions of the dips in 
Fig.~\ref{CPR}(a). %
The origin of the dips can be further confirmed by exploring the CPR for
different frequencies, as we do in Fig.~\ref{CPR}(b). Here, one can see
that by decreasing the frequency, the dip of order $n=1$ moves to higher 
values of $\varphi$ and disappears for $\hbar 
\omega \leq 0.4\Delta$,  in agreement with Eq.~(\ref{phin}). 

One can gain further insight by analyzing this problem in terms of
a TLM  that describes the dynamics of our QPC in
terms of the ABSs \cite{Ivanov1999,Zazunov2003}. We  
consider the TLM of Ref. \cite{Zazunov2003}, whose effective Hamiltonian
 in the instantaneous basis of ABSs reads %\cite{note1}
\begin{equation}
\hat H_{\rm A}(t) = E^+_{\rm A}(\phi(t))\hat{\sigma}_z -
\frac{r \tau \Delta^2 \sin^2(\phi(t)/2)} {4[E^+_{\rm A}(\phi(t))]^2}
\hbar \dot{\phi}(t) \hat{\sigma}_y , \label{tlm}
\end{equation}
where $\hat{\sigma}_{y,z}$ are Pauli matrices, $r= \sqrt{1-\tau}$ and
$\dot{\phi}(t) = \partial \phi(t) / \partial t$. We have computed the
CPR from this model using a Floquet approach \cite{note2}. In
Fig.~\ref{CPR}(a) we show a comparison of the results of this TLM
with the exact results. There is an excellent agreement
in a wide range of phases and in particular, the TLM is able to
reproduce the current dips. A discrepancy occurs at phases
close to $\pi$, which is understandable as the model assumes that
$\hbar\dot{\phi}(t)\sim\alpha\omega\ll2E_{\rm A}^{+}$, while 
for $\varphi \sim \pi$ and high $\tau$, the ABSs are very close to each
other and the assumption no longer holds. Using a rotating wave type
approximation  we also
obtain from the TLM the following analytical expression for the DC current at the
first two resonances,
\begin{equation}
I(\varphi,\omega,\alpha) \simeq \frac{2eE_A'}{\hbar} \left(1 - \frac{\gamma_1^2}
{\delta_1^2 + \gamma_1^2} \right) \left(1 - \frac{\gamma_2^2}
{\delta_2^2 + \gamma_2^2} \right) .\label{2dips}
\end{equation}
Here and below, $E_{\rm A}'$ and $E_{\rm A}''$ are the first and second 
derivatives of $E^+_{\rm A}$ with respect to the phase. The detunings 
$\delta_{1,2}$ are given by $\delta_1 = E^+_{\rm A}(\varphi) - \hbar \omega/2
+ \alpha^2 E_{\rm A}''(\varphi) + 3 r^2 \alpha ^2 \hbar\omega
(\Delta^2-E_A^+(\varphi)^2)^2/(32E_A^+(\varphi)^4)$, and $\delta_2 
= E^+_{\rm A}(\varphi) - \hbar \omega + \alpha^2 E^{\prime \prime}_{\rm A}
(\varphi) - r ^2\alpha^2 \hbar \omega (\Delta^2-E^+_{\rm A}(\varphi)^2)^2/
(12E^+_{\rm A}(\varphi)^4)$, and include shifts in the ABS energies caused 
by the microwave field. The resonance widths are $\gamma_1 = r 
\alpha \hbar \omega (\Delta^2 - E^+_{\rm A}(\varphi)^2)/ (4E^+(\varphi)^2)$ 
and $\gamma_2 = r \alpha^2 \hbar \omega E^{\prime}_{\rm A}(\varphi)
(2\Delta^2 - E^+_{\rm A}(\varphi)^2)/(2E^+_{\rm A}(\varphi)^3)$,
in general being proportional to the coupling strength, $\gamma_n\propto\alpha^n$.
For small power, Eq.~(\ref{2dips}) is in a good agreement with the exact 
solution for the current-phase relation. We also find that within the TLM 
the current vanishes completely at the resonances 
(given by the condition $\delta_n = 0$), as a consequence of the fact that at these points the long-time average populations on the ABSs are equal, so that the currents carried by the two states cancel exactly
%(for  $n=1$ and $\alpha \ll 1$ this cancellation was 
%first predicted  in Ref. \cite{Shumeiko1993}).
\cite{note3}. Transitions from the continuum, however, make the current at the resonances finite and dependent on the frequency and the excitation energies. This is seen for example in the exact result for the second dip shown in Fig.~\ref{CPR} \cite{note4}.
\begin{figure*}[t]
\begin{center}
\includegraphics[width=0.75\textwidth]{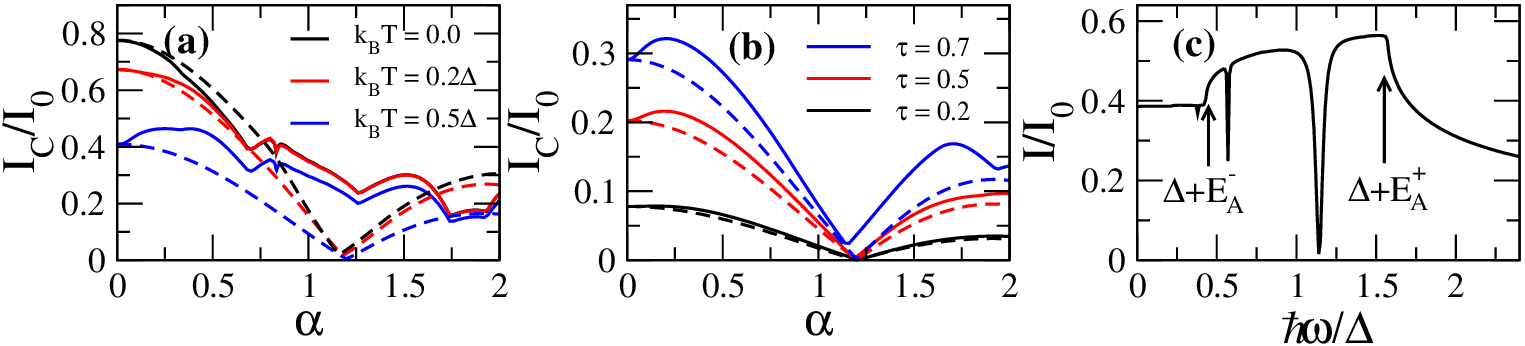}
\caption{(Color online) Panels (a-b) show the critical current as a function of 
$\alpha$ for (a) $\hbar \omega = 0.3\Delta$, $\tau=0.95$ and different values of
$T$, and (b) $\hbar\omega=0.3\Delta$, $k_{\rm B}T = 0.5\Delta$ and different
values of $\tau$. In both panels the solid lines correspond to exact results
and the dashed ones to Eq.~(\ref{Iad}). (c)
Current as a function of the frequency for $\varphi = 2.0$, $k_{\rm B}T = 0.5 \Delta$, $\tau=0.95$ and $\alpha = 0.1$.} \label{critical}
\end{center}
\end{figure*}

Let us now discuss the dependence of the CPR on the radiation power. In
Fig.~\ref{CPR}(c) we show the CPR for $\tau=0.95$, $\hbar \omega =
0.3 \Delta$ and different values of $\alpha$. As $\alpha$ increases, the
CPR is drastically modified and the current is not only strongly suppressed
around the phase values given by Eq.~(\ref{phin}), but everywhere. Notice
also that in certain regions, particularly at large phases, the current even
reverses its sign. It is also worth stressing that as $\alpha$ increases,
we find larger deviations between the exact results and those of the TLM
(not shown here) due to multi-photon processes connecting the ABSs and the
continuum of states. For the sake of completeness, we illustrate in Fig.~\ref{CPR}(d) the influence 
of the transmission in the CPR for $\hbar \omega = 0.3 \Delta$ and $\alpha = 0.1$. 
In this case, for $\tau \lesssim 0.9$ the CPR can be accurately described 
with the adiabatic approximation for all phases since the transitions
between the ABSs are very unlikely.

We turn now to the analysis of the critical current $I_C$, i.e.\ the maximum 
value of the DC Josephson current. In Fig.~\ref{critical}(a) we show the 
critical current as a function of $\alpha$ for several temperatures. One clearly sees that the adiabatic 
approximation (dashed lines) only describes correctly the behavior of $I_C$ 
when the frequency, power and temperature are low enough so that the microwaves 
cannot induce transitions between the ABSs and between them and the continuum.
The most striking result in Fig.~\ref{critical}(a) is the enhancement of $I_C$ with respect to the
case $\alpha=0$  as the temperature is raised.  In Fig.~\ref{critical}(b) we show that 
this occurs in a wide range of transmission values. The enhancement of the 
critical current by a microwave field has been predicted and observed in 
superconducting microbridges \cite{Wyatt1966,Eliashberg1970} and in proximity 
effect structures \cite{Notarys1973,Virtanen2010}. In both cases the enhancement
is due to a redistribution of the excitations induced by the field. In our 
case the underlying mechanism is similar. At finite temperature the lower ABS 
is not fully occupied and the microwave field can promote quasiparticles from 
the continuum to this state if \ $\hbar \omega \ge \Delta + E^-_{\rm A}(\varphi) $, for 
values of $\varphi$  close to that where the maximum takes place (see 
process \emph{c} in Fig.~\ref{model}). 
This naturally results in an enhancement of the supercurrent. We illustrate this argument in 
Fig.~\ref{critical}(c), where we show the frequency dependence of the 
supercurrent for a fixed value of the phase $\varphi_0 = 2.0$. As one can see, 
the current remains largely unaffected until $\hbar \omega$ reaches $\Delta + 
E^-_{\rm A}(\varphi_0) \approx 0.46 \Delta$, where it starts to increase. 
Then, as the frequency increases further one can observe the appearance of 
two dips, corresponding to the resonances at $\hbar\omega=E^+_{\rm A}/2$ and 
$E^+_{\rm A}$. For larger values of $\hbar\omega$ one finally sees a decrease 
of the current at $\Delta + E^+_{\rm A}(\varphi_0) \approx 1.54\Delta$ due to
processes that promote quasiparticles from the continuum to the upper ABS
(see process \emph{d} in Fig.~\ref{model}). Due to electron-hole symmetry similar 
transitions exist between the ABS and the upper continuum. Obviously, the 
enhanced supercurrent cannot be explained in terms of any TLM since 
it involves the continuum of states.

In summary, we present here a microscopic theory of the microwave-assisted
supercurrent in quantum point contacts. It predicts the appearance of a
variety of novel phenomena that in general are out of the scope of simple 
approximations and two-level models. Our results are of relevance for many
different types of weak links and in particular, they can be quantitatively 
tested in the context of atomic contacts. 

We thank A. Levy Yeyati, C. Urbina, M. Feigelman and C. Tejedor for motivating discussions.
This work was supported by the Spanish MICINN (contract FIS2008-04209), EC 
funded ULTI Project Transnational Access in Programme FP6 (Contract 
RITA-CT-2003-505313). T.T.H.\ acknowledges the funding by the Academy
of Finland and the ERC (Grant No. 240362-Heattronics).

%%%%%%%%%%%%%%%%%%%%%%%%%%%%%%%%%%%%


\begin{thebibliography}{10}
\bibitem{note2} 
Details of the calculation will be published elsewhere.
\bibitem{note4} 
With the parameters of Fig.~\ref{CPR}, the dip obtained from the TLM for $n=3$  is not obtained from the exact model.
\end{thebibliography}
\end{document}